# Evidence of Negative Heat Capacity, Rigidity Percolation and Intermediate Phase in Fast Ion Conducting Conditional Glass Forming System

B. Tanujit and S. Asokan[a]

Department of Instrumentation and Applied Physics
Indian Institute of Science, Bangalore – 560 012, India

**Abstract**

In this work, we observe the rigidity percolation phenomena in a fast ion conducting, conditional glass forming system $(AgI)_{75-x}$-$(Ag_2O)_{25}$-$(MoO_3)_x$. To find out where, why and how the rigidity percolation phenomenon occurs within the range of $20 \leq x \leq 37.5$, calorimetry and photoelectron spectroscopy experiments are performed. The temperature dependence of heat capacity (normalized) at glass transition temperature ($T_g$), exhibits fluctuations for samples with higher AgI concentration. This specific quality attributes to fragile glass. The wide range of composition accommodates both the fragile and strong glasses, and therefore a fragility threshold. The heat capacity (absolute) values, at $T_g$ when plotted over the whole range of compositions, exhibits an abrupt sign shift, from negative to positive, revealing the fragility threshold. The appearance of negative heat capacity has been corroborated with the thermodynamic behavior of nanoclusters. This technique has been identified as a novel method to recognize the existence of nanoclusters in this type of glasses. The photoelectron spectroscopy study shows the formation of essential covalent structural units, [– Mo – O – Ag – O –] and complex molybdenum oxides in the positive heat capacity region. Finally, the non-reversing enthalpy profile has been studied over the whole composition range. The global, square well minima sandwiched between floppy and stress rigid region has been identified to be the intermediate phase, within the range $32.25 \leq x \leq 35$.

[a] Corresponding Author email: sundarrajan.asokan@gmail.com, sasokan@iisc.ac.in

Fax: 91-80-23608686; Phone. +91-80-22933195, 91-80-22932271

## I. Introduction

Kauzmann's glass condition states that the quantity $\frac{\partial (H_{liquid} - H_{crystal})}{\partial T}$ drops to zero during the glass transition [1], where H signifies enthalpy of the indexed state and T is the temperature. In other words, across $T_g$, the configurational entropy ($S_c$) remains continuous during a liquid to solid or supercooled liquid to glass transition [1] i.e.,

$$\Delta S_c(T_g) = \left(S_{c.liquid}(T_g) - S_{c.solid}(T_g)\right) = 0 \tag{1}$$

To realize this phenomenon microscopically, J. C. Phillips and M. F. Thorpe [2-5] introduced a 'site-bond' type model, for systems that constitute covalent bonds, dominantly. This bond specification considers (i) nearest neighbor, central, two-body bond-stretching and (ii) non-central, three-body bond-bending forces. The glass condition in equation 1 convolutes into a constraint equation when associated with this site-bond model, average first coordination number and medium range order. The constraint equation is given as,

$$N_{Con} = N_d \tag{2}$$

Where, $N_{con}$ is the number of interatomic force field constraints per atom and $N_d$ is the number of vector degrees of freedom per atom [2, 5]. In a covalent network structure, this condition (equation 2) behaves as the rigidity percolation threshold. The condition $N_{Con} < N_d$ refers to a floppy, polymeric phase, for instance a linear network in three dimension i.e. $N_{Con} = 2$ and $N_d = 3$. As the constraints get accommodated more within the system, with compositional variation, the interconnectivity among the sites increases and rigidity percolates through the system; this phase is known as stressed rigid phase when $N_{Con} > N_d$ [5].

Contemporaneous study on the phenomena of photo-melting in $As_2S_3$ glass [6] and giant Photocontraction in obliquely deposited porous $GeSe_2$ [7] identified another stable phase where the network is mechanically rigid but stress free [8]. This 'trapped in limbo' phase in rigidity transition corresponds to equation 2 [9]. Boolchand et al. [8-11] discovered this stress-free, intermediate phase (IP) in many other glass systems. IP glasses exhibits non-aging behavior, optimal glass forming tendency [12] and it possess self-organization functionality [9] which is the global reconnection of chemical bonds to form stress-free networks [12]. This functionality along with the Phillips- Thorpe model has been used to understand protein folding and protein phase transition [13]. Self-organization and IP has been recognized in thin-film transistor (TFT) used in liquid crystal display (LCD) [14]; and further exploited in various disciplines e.g. Soft condensed matter, Computer science, Electrical engineering, Protein science etc. [12, 14].

Early studies on rigidity transition and IP [8-11] were mainly conducted on chalcogen (S, Se, Te) based glasses that attribute a covalent, continuous random network (CRN) topology which is essential for rigidity transition. Oxygen, on the other hand, having high electronegativity and a tendency to form $O^-$ and $O^{2-}$, differs significantly from other chalcogens [15] in terms of network topology and chemical properties. Moreover, when network modifier oxide is introduced into a base oxide glass, non-bridging oxygen (NBO) forms and ionically bonds with modifier cation; this phenomenon alters the network topology significantly [15]. Increasing modifier oxide concentration causes systematic degradation of the network topology i.e. de-polymerization of CRN and modification in $Q_n$ species takes place as ultra ($Q_3$) → meta ($Q_2$) → pyro ($Q_1$) → ortho ($Q_0$), where $Q_n = [AO_{\frac{n}{2}}O_m]^{m-}$, $(n + m) \leq 4$ and A = P, Ge and Si [20]. But interestingly, three phased rigidity transition has been observed in some oxide glasses (Type I) e.g. AgI based fast ion conducting (FIC) glass, $(AgI)_x(AgPO_3)_{1-x}$ [16-17], alkali germanate glass, $(Na_2O)_x(GeO_2)_{1-x}$ [18], alkali silicate glass, $(Na_2O)_x(SiO_2)_{1-x}$ [19]. Although, the constituent glass forming agent for these glasses are $P_2O_5$, $GeO_2$ and $SiO_2$ respectively, all of which are strongly covalent glass formers [20-21] and follow Zachariasen's rules for glass formation [20]. However, heavy, transition metals are capable of existing in multiple valance states and known to be 'conditional' oxide glass former [20]. Pure liquids of these transition metal oxide compounds essentially require modifier or another network forming oxides for glass formation, BO → NBO conversion, different $Q_n$ species and modifier cation formation. The ionicity these metallic elements e.g. Mo, W and ionic bonding or 'partial covalence' [22-23] between modifier cations and NBOs, significantly enhance the non-covalent nature of these type of glassy systems (Type II). De-polymerization affects the covalent graph and form fragile molecular solid where interatomic strong and intermolecular weak force coexist [15]. Fragile glasses that lack directional bonds [24] exhibit (i) deviation from Arrhenius type behavior in the viscosity-temperature profile (ii) fluctuation in the heat capacity profile at $T_g$. Moreover, structural disparity, in terms of the origin of First Sharp Diffraction Peak (FSDP), between these two types of glasses (Type I and Type II) reported and argued by Swenson et al. [25] in their Neutron Diffraction study. In case of molecular (oxide) glasses (Type II), FSDP is significantly contributed by oxygen which is in contrary with Type I glass formers e.g. phosphate [26] and borate [27] glasses. Thus, these conditional glass former, transition metal oxides are significantly different, in a chemical and topological sense, from CRN forming Chalcogenide and metalloid/non-metal oxides.

In this present work, we report rigidity percolation phenomena in $(AgI)_{75-x}$-$(Ag_2O)_{25}$-$(MoO_3)_x$ fast ion conducting glassy system. Within a broad composition range of $20 \leq x \leq 37.5$, where, why and how the rigidity transition should occur? To answer these questions, Alternating differential scanning calorimetry (ADSC/MTDSC) and X-ray photoelectron spectroscopy (XPS) experiments are conducted. Being a conditional glass system, the wide range of composition accommodates two classes. This classification is based on fragility and these two different classes are 'fragile' and 'strong'. Experimentally, a contrast between these two classes can be established from the fluctuation behavior of heat capacity near $T_g$. The 'margin of contrast' is known as the fragility threshold. The fragile region consists of glasses with more ionicity than the glasses from strong region. Moreover, the rigidity percolation phenomenon should happen in this strong region. Detailed study on heat capacity provides striking results. The heat capacity (normalized) profile confirms the fluctuating behavior but remains incapable to identify the fragility threshold. This is done by observing a sharp sign shift, from negative to positive in heat capacity (absolute) value, at $T_g$ while plotted against compositions. Furthermore, the negative heat capacity is attributed to the thermodynamic behavior of nanoclusters. And this has been corroborated to the presence of nanocluster in the glasses that comprises the fragile region. The ionic bonds in this region gradually get replaced by the covalent while increasing $MoO_3$ concentration and form the strong region. The XPS study recognizes the bond as [– Mo – (O – Ag) – $O_L$ –] and complex molybdenum oxide in the covalent region. Finally, the $\Delta H_{nr}$ value exhibits a global minima representing the IP, sandwiched between under-constrained floppy phases (FP) and over-constrained stressed rigid phase (SRP).

## II. Experimental
### A. Glass preparation

$(AgI)_{75-x}$-$(Ag_2O)_{25}$-$(MoO_3)_x$ solid electrolyte glasses, within the composition region $20 \leq x \leq 37.5$, have been synthesized, by thoroughly mixing constituent compounds as fine homogeneous powder, and melting it in a 2450 MHz-900 watts microwave oven for 10-12 minutes, before quenching the melt, down to room temperature between two steel plates. The idea of keeping $Ag_2O$ concentration constant while varying AgI and $MoO_3$ concentration, along with the novelty of using microwave heating technique has been discussed in detail elsewhere. [28]

### B. Material Characterizations and Analysis

ADSC study are carried out using a Mettler Toledo 822e ADSC instrument operated at 3$^o$C/min scan rate with 1$^o$C scan amplitude, ranging from 40$^o$C to 200$^o$C for 'as quenched' samples ( ~ 15 mg) of approximately equal thickness. The obtained heat flow and heat capacity data has been analyzed using the

STAR$^e$ and OriginPro software. The X-ray photoelectron spectroscopy (XPS) study has been carried out in an AXIS Ultra XPS instrument with monochromatic Al X-ray source. The binding energy scale has been calibrated with C 1s 284.8 eV peak. The obtained data are fitted with Gaussian-Lorentzian peak profile after U 2 Tougaard background subtraction with CasaXPS software.

### III. Results and Discussions
#### A. Fragility Threshold and Negative Heat Capacity

ADSC experimentation involves a sinusoidal modulated heat flow response and the resultant heat flow gets associated with a non-zero phase lag (φ). But other than melting, this phase correction of heat capacity is negligible [29]. Thus, in this work, in-phase heat capacity is considered to be the measure for the heat capacity. Firstly, large fluctuation in heat capacity near $T_g$ is a manifestation for deviation from Arrhenius type relaxation behavior due to increase in non-directional interatomic / intermolecular bonds; whereas significant amount of directional covalent bonds form strong glass system [30, 31] which is a necessary requirement for rigidity transition. In earlier works [30, 32], this fluctuation has been reported to have a shape of 'overshooting' near $T_g$, during heating; and during cooling this fluctuation is absent, leaving a trace of hysteresis. In the present study, the fluctuation is determined by any sharp change in the tangential direction. Figure 1(a) shows $C_p$ (normalized) versus $T/T_g$ plot for two different representative samples, $(AgI)_{50}$-$(Ag_2O)_{25}$-$(MoO_3)_{25}$ (AIAMo-5) with $T_g$ = 73.39 $^o$C and $(AgI)_{42.25}$-$(Ag_2O)_{25}$-$(MoO_3)_{32.75}$ (AIAMo-15) with $T_g$ = 109.39 $^o$C. Clearly, near $T_g$, glass with higher AgI concentration i.e. AIAMo-5 exhibits more fluctuation in $C_p$-$T/T_g$ profile than the other. Sharp change in tangential direction happens at point A and B (Figure 1(a)) in this glass, which is absent in AIAMo-15. The present result corroborates with the understanding of the role of AgI. The role of AgI is to expand the free volume to enhance the ionic conductivity [33]. The expansion in glass matrix causes creation of free volume and lower coordination that directly impact the $C_p$-$T/T_g$ profile; with higher AgI concentration, the glass system becomes more fragile.

This nature of fluctuation in the $C_p$-$T/T_g$ profile eventually solves two different problems. Firstly, the thermodynamic relation [34],

$$C_p = T \left(\frac{\partial S}{\partial T}\right)_p \tag{3}$$

Where, S is entropy, can be represented as,

$$\int_{S'(T_1)}^{S''(T_2)} dS = \int_{T_1}^{T_2} C_p \frac{T_g}{T} \frac{dT}{T_g} \tag{4}$$

Where, $S'(T_1)$ and $S''(T_2)$ are the entropies corresponding to temperature $T_1$ and $T_2$. These two temperatures, $T_1$ and $T_2$ are basically the start and end temperature of ADSC scan. Figure 1(b) presents this integration of obtained data which are presented in Figure 1(a). The integrations for these discrete data are done using the trapezoidal method.

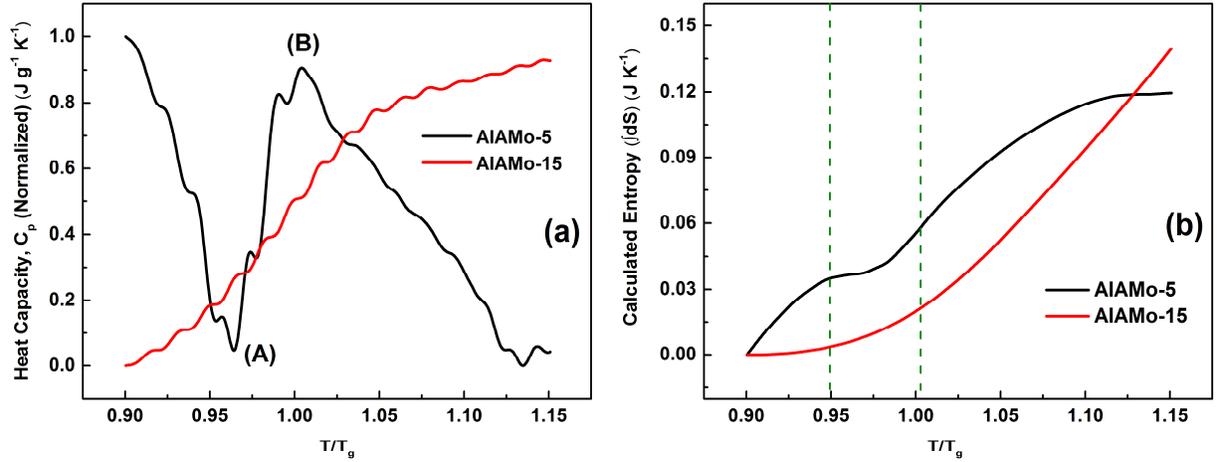

*Figure 1.(a) Heat capacity, $C_p$ ($Jg^{-1}K^{-1}$) (normalized) versus $T/T_g$ plot for two representative samples, $(AgI)_{50}$-$(Ag_2O)_{25}$-$(MoO_3)_{25}$ (AIAMo-5) and $(AgI)_{42.25}$-$(Ag_2O)_{25}$-$(MoO_3)_{32.75}$ (AIAMo-15). (b) Calculated entropy values from equation-4 and heat capacity data for samples AIAMo-5 and AIAMo-15*

The entropy curve in Figure 1(b) exhibits a dent or an inverted curvature within two green marked lines. This phenomenon has been observer in the study of cluster formation of 147 sodium atoms [35].This dent has been discussed in this [35] work, results in a negative slope in the micro-canonical caloric curve which means the corresponding heat capacity becomes negative. Moreover, micro-canonical caloric curve involves measurement of temperature of an isolated cluster that has been formed within this system. This implicates cluster formation within the present system.

Secondly, the trend of the $C_p$-$T/T_g$ profile over the whole composition range suggests that the nature of fluctuation systematically varies with composition and it is possible to identify a threshold. Interestingly, a change appears abruptly in the sign of heat capacity (absolute) values. As the sign of $C_p$ for an individual sample remains same over the ADSC scanning range (40°C-200°C), we consider $C_p$ at $T_g$ i.e. $C_p(T_g)$ as an indicator of the sign of $C_p$. Figure 2(a) shows $C_p(T_g)$ values for all the compositions along with $T_g$. The $T_g$ increases monotonically with the increase in $MoO_3$ concentration. But $C_p(T_g)$ shifts from negative (Region I) to positive (Region II) almost abruptly at a threshold between $(AgI)_{30}$-$(Ag_2O)_{25}$-$(MoO_3)_{45}$ and $(AgI)_{31.25}$-$(Ag_2O)_{25}$-$(MoO_3)_{43.75}$. This shift in $C_p(T_g)$ profile infers very important conclusion about the structure of the sample. Figure 2(b) represents the heat capacity (Absolute) profile of with sample temperature. The sharp fall of $C_p$ for AlAMo-15, near melting must have caused by incongruence and heterogeneous melting. Samples with negative $C_p(T_g)$ value, exhibit the fluctuations in normalized $C_p$-$T/T_g$ profile than the samples with positive $C_p(T_g)$ value. This scenario has to be discussed from two perspectives; what does negative $C_p$ mean thermodynamically and exclusively for this sample.

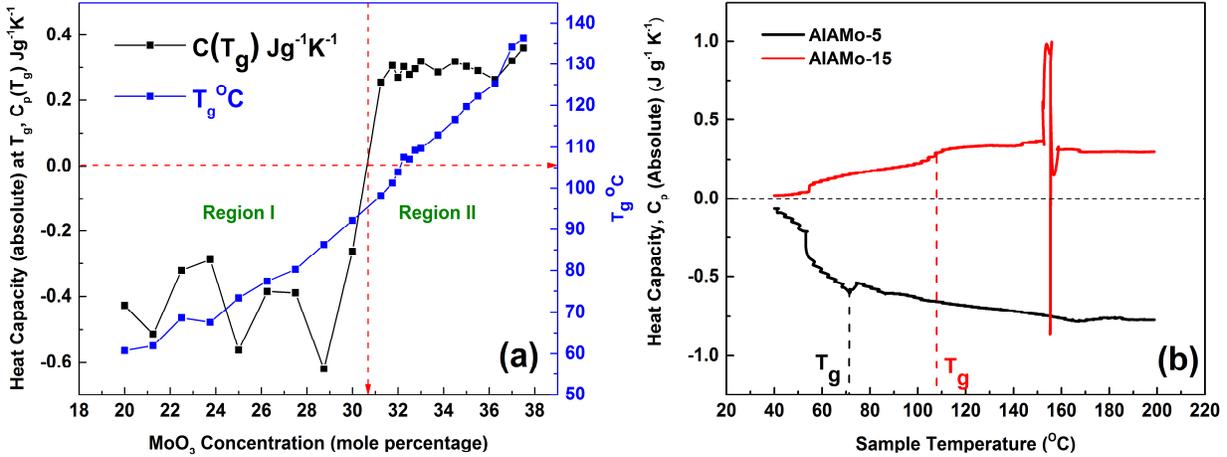

*Figure 2: (a) Heat capacity (absolute) values at $T_g$, $C_p(T_g)$ ($Jg^{-1}K^{-1}$) and glass transition temperature, $T_g \,^oC$ vs. $MoO_3$ concentration (mole percentage)(b) Heat capacity, $C_p$ ($Jg^{-1}K^{-1}$) (Absolute) versus sample temperature ($^oC$) plot for two representative samples, $(AgI)_{50}$-$(Ag_2O)_{25}$-$(MoO_3)_{25}$ (AlAMo-5) and $(AgI)_{42.25}$-$(Ag_2O)_{25}$-$(MoO_3)_{32.75}$ (AlAMo-15).*

In the thermodynamic context, negative $C_V$ is a compelling topic in the branch that concerns thermodynamic behavior of nanoclusters where two different structural phases separated by free energy barrier. Negative $C_V$ ($\equiv (\partial E/\partial T)_V$) is a consequence of the 'S' shaped bend in the caloric curve, which has been argued to be the indication of a dynamic phase coexistence. At a critical temperature, small finite system undergoes critical oscillation between two metastable states, giving rise to the negative slope in the caloric curve [36-41]. Noticeably, these studies on nanoclusters consider the heat capacity at constant volume i.e. $C_V$, not $C_p$. Besides, $C_p > C_V$ [36] (in general, equation 7) i.e. if $C_p$ is negative, $C_V$ must be negative, but not otherwise. This eventually confirms the existence of nanoclusters in the present sample, within a compositional threshold. Still, we tried to find traces of equivalence between these two heat capacities that will correlate the structure of the solid and this thermodynamic aspect, because in a solid with supercooled liquid structure, this inequality relation might not hold.

In an important study on the 'effect of pressure on conductivity in $AgI$-$Ag_2O$-$MoO_3$ glassy system' by H Senapati et al. [42] suggested a tissue and cluster type structural model where the whole glass structure is composed of clusters those are connected by less dense and compressible tissue material. Applied pressure influences the tissue and molds it into a structure similar to that of the cluster.

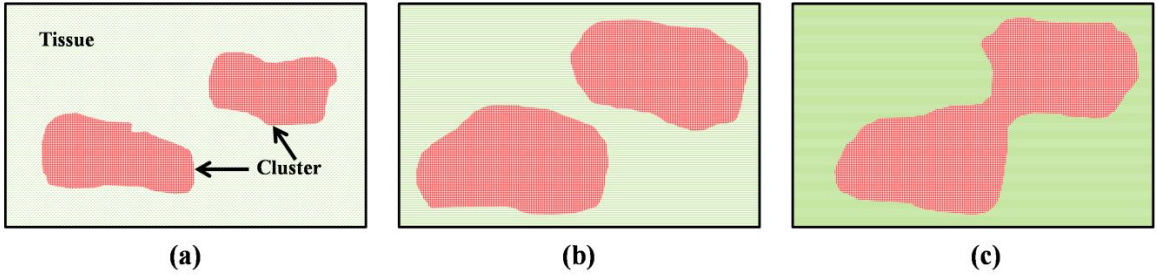

*Figure 3: (a) A segment of $V_{Total}$ which is under pressure constitutes of dense cluster and less dense, compressible tissue. (b) As the pressure increases, the tissue reorganizes to yield a structure similar to the cluster. (c) Higher pressure can change tissue region in between two clusters and thus connecting them while increasing the cluster volume. Pressure increases from (a) to (c)*

Schematics in Figure 3 show the effect of pressure on the total volume ($V_{Total}$) that constitutes of tissue and cluster. In other words, with pressure (p), the tissue volume ($V_{tissue}$) decreases in expense of the increase in cluster volume ($V_{cluster}$) i.e. in an isothermal process, $V_{Total}$ under pressure remain constant;

$$V_{Total} = V_{tissue} + V_{cluster} = Const. \tag{5}$$

Thus the thermodynamic quantity,

$$\left(\frac{\partial P}{\partial V}\right)_T = 0 \tag{6}$$

Where T is temperature; now the thermodynamic relation,

$$C_p - C_V = -\left(\frac{\partial P}{\partial V}\right)_T T \left(\frac{\partial V}{\partial T}\right)_P^2 \tag{7}$$

Where $C_p$ and $C_V$ are isobaric and isochoric heat capacities respectively, becomes

$$C_p - C_V = 0 \tag{8}$$

Hence the experimentally obtained $C_p$ and the relation (8) substantiate an equal sign for $C_V$, always.

Thus, incorporating these two assertions, sample based and thermodynamic, we suggest that experimentally obtained negative $C_p$ is a reflection of a structural phase consisted of tissue and cluster discussed earlier. Within this region, samples exhibit the presence of nanocluster, negative heat capacity, non-Arrhenius relaxation, loss of network connectivity and hence more fragility. Thus this fragility threshold, defined as the sign shift of $C_p$, can be regarded as a shift from a less connected to a highly connected network. To explain this nature of network connectivity or bonding in terms of chemical states of the constituents, we performed XPS for some of the samples from region I and II.

### B. XPS Study

High resolution XPS peaks for Ag 3d, Mo 3d and O 1s spectrums have been obtained. The binding energy (BE) for Ag $3d_{5/2}$ peak is at 367.3 eV for region I and 366.8 eV for region II i.e. $\Delta BE = BE$ (final) $- BE$ (initial) $< 0$. Early studies has assigned the 367.3 eV peak to Ag $3d_{5/2}$ peak for $Ag_2O$ where the oxidation state of Ag is +1 and 366.8 eV peak for AgO that contains an equimolar mixture of Ag (+1) and Ag (+3) oxidation states that appears to have an average valency of $Ag^{2+}$ [43, 44]. Surface content of silver for the sample mostly gets oxidized, giving rise to only Ag $3d_{5/2}$ peak for Ag-O bond, whilst Ag-I remains very weak, indicating that the features of AgI comes within the bulk property of the sample. The binding energy for Mo $3d_{5/2}$ in the spectrum appears at 230.63 eV and 231.18 eV. This suggests that the oxidation of Mo consisted of $Mo^{4+}$ and $Mo^{5+}$ species respectively [45, 46]. The relative intensities (in percentage) for peaks of both the Ag $3d_{5/2}$ and Mo $3d_{5/2}$ remain unchanged over the whole composition range.

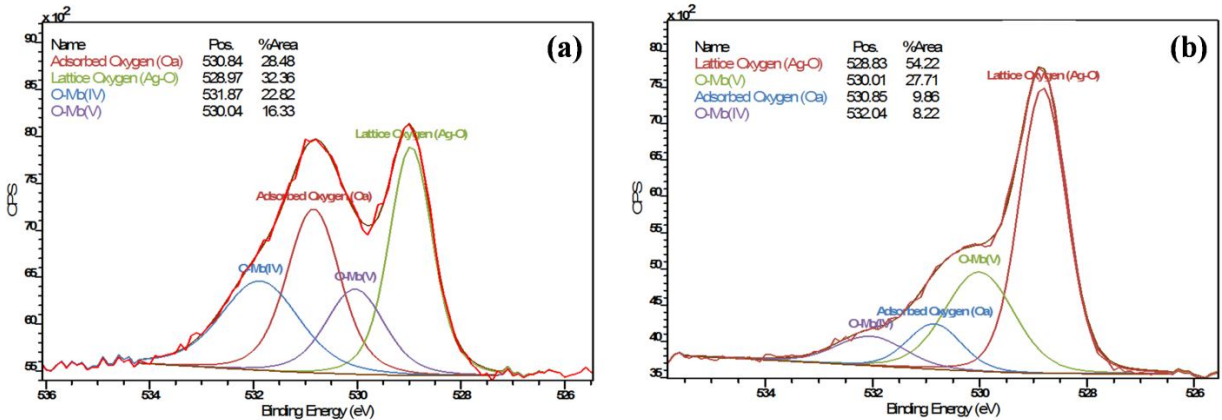

*Figure 4: O 1s spectrum for representative samples (a) For $(AgI)_{47.5}$-$(Ag_2O)_{25}$-$(MoO_3)_{26.25}$ (AIAMo-6) which belongs to Region-I (b) For $(AgI)_{42.5}$-$(Ag_2O)_{25}$-$(MoO_3)_{32.5}$ (AIAMo-14)*

Figure 4 shows deconvoluted O 1s spectrum for representative samples from region I and II with composition (a) $(AgI)_{47.5}$-$(Ag_2O)_{25}$-$(MoO_3)_{26.25}$ (AIAMo-6) and (b) $(AgI)_{42.5}$-$(Ag_2O)_{25}$-$(MoO_3)_{32.5}$ (AIAMo-14) respectively. The O 1s spectrum consists of lattice oxygen ($O_L$) that is related to Ag-O bonding, adsorbed oxygen ($O_A$) that contains hydroxyl group, lattice defects and surface organic contaminations that result in C 1s spectrum, oxygen bonded with $Mo^{4+}$ ($O_{Mo(IV)}$) and $Mo^{5+}$ ($O_{Mo(V)}$). The peak at 528.8 eV has been assigned to $O_L$, 530.7 eV to $O_A$ [47]. 531.7 eV to $O_{Mo(IV)}$ [48] and 530.0 eV to $O_{Mo(V)}$.

Figure 5 shows the change in relative intensities of all the oxygen peaks over the composition range. The relative intensity increases rapidly for $O_L$ from region-I to II in expense of other oxygen species.

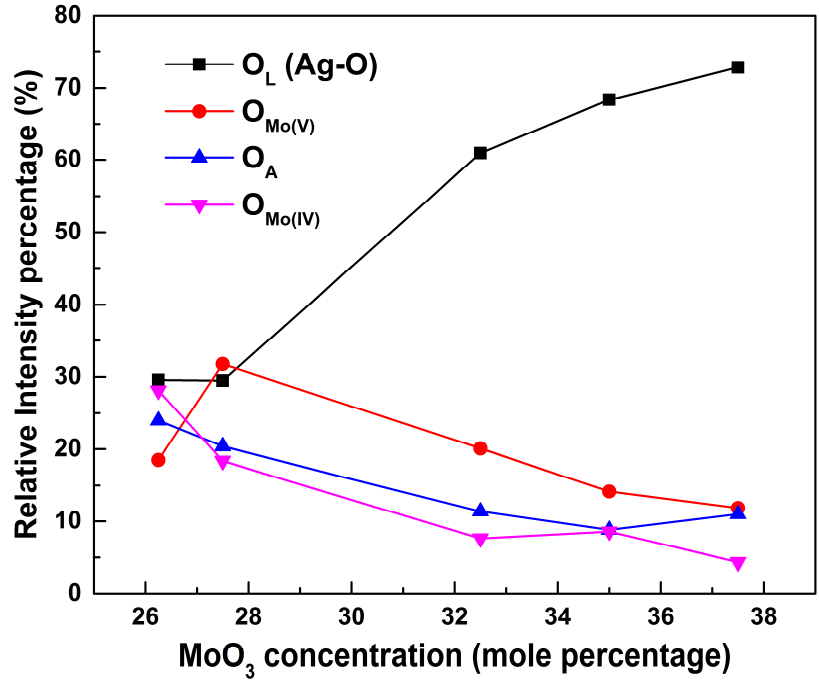

*Figure 5: Relative intensities (in percentage) of oxygen species present in the sample*

The adsorbed oxygen intensity decrement and its contribution to lattice oxygen intensity is interpreted by the equation

$$O_2(adsorbed) + 4e^- \rightarrow 2O^{2-}\ (lattice) \qquad (9)$$

Part of these excess lattice oxide populates around Ag while eventually increasing its oxidation state; this situation gets associated with the shift of Ag $3d_{5/2}$ peak, a transition from Ag(I)-O in region-I to Ag(I, III)-O in region-II. $O^{2-}$ is nucleophilic with a tendency to form metal-oxygen-metal bonds [49] and hence, part of $O_L$ gets associated with less electronegative Mo and consequently modifies the O-Mo bonds. Both $O_{Mo(IV)}$ and $O_{Mo(V)}$ shifts towards lower binding energy (~ 0.21 eV for $O_{Mo(IV)}$ and ~ 0.51 eV for $O_{Mo(V)}$ ) with a significant decrease in relative intensity. On the other hand, $Mo^{4+}$ and $Mo^{5+}$ peak position and intensity remains unchanged throughout the composition range. Thus, the oxidation of Ag and subsequent reduction of $O_{Mo(IV)}$ and $O_{Mo(V)}$ doesn't affect $O_L$ and Mo species respectively. This situation suggest electron sharing and hence bond formation that can be interpreted as [– Mo – (O – Ag) – $O_L$ –]. The oxidation states of Mo and $O_L$ doesn't get affected in expense of the formation the (O–Ag) bond. [– Mo – (O – Ag) – $O_L$ –] is the structural unit of this glass that properly forms in region II. These units are either absent or partially formed in region I that might cause the clustering and hence negative heat capacity. Besides, the decrease in relative intensity of $O_{Mo(IV)}$ and $O_{Mo(V)}$ suggests that more Mo is being shared with less O that indicates formation of complex oxides of Mo. These conclusions together suggest an overall increase in covalent bonds and hence bridging oxygen (BO) during the transition from region I to region II. Previous discussion on negative heat capacity together with the present one concludes where and why the rigidity percolation should occur in this ion conducting conditional glass forming system.

### C. $\Delta H_{nr}$ and Intermediate Phase

The change in configurational entropy during liquid to solid transition presented in equation 1, is directly related to two measurable quantities, non-reversing enthalpy ($\Delta H_{nr}$) and $T_g$ as [12]

$$\Delta S_c = \Delta H_{nr}/T_g \tag{10}$$

When the model solid is glass, $\Delta S_c = 0$ which infers that configurational freedom of glass and liquid are same; otherwise $\Delta S_c > 0$. Apparently, this feature of configurational entropy is directly reflected upon $\Delta H_{nr}$. The ADSC experiment involves a sinusoidal temperature modulation, represented as,

$$T = T_0 + \beta t + B \sin(\omega t) \tag{11}$$

Where, T is the temperature, $T_0$ is the initial heat bath temperature, $\beta$ is the heating rate, t is time, B is the amplitude of modulation and $\omega$ is the modulation frequency. The heat flow rate is given by,

$$\frac{dQ}{dt} = C_p \frac{dT}{dt} + f(t,T) \tag{12}$$

Where, dQ/dt is the heat flow rate, $C_p$ represents the heat capacity and f(t, T) represents some function of time and temperature that is associated with responses due to physical and chemical transformation. This temperature modulation can separately identify two different types of thermal events: first, those of which respond instantaneously to any change in the modulation, either amplitude (B) or frequency ($\omega$), result in the reversing heat flow i.e. ($C_p$ dT/dt) in equation 12; e.g. glass transition. Second, those of which exhibit sluggish response to ramp modulation, which is observed in non-reversing heat flow component i.e. f(t, T), for instance enthalpy relaxation at the glass transition [50]. Finally, the non-reversing heat flow signal exhibits a Gaussian profile encompassing $T_g$, the area under that profile yields the value of $\Delta H_{nr}$ [17].

Figure 6 shows the non-reversing enthalpy profile over the whole composition range, $\Delta H_{nr}$ ($Jg^{-1}$) – x profile. The data acquisition technique has been rigorously discussed in earlier works by Micoulaut et al. [17].

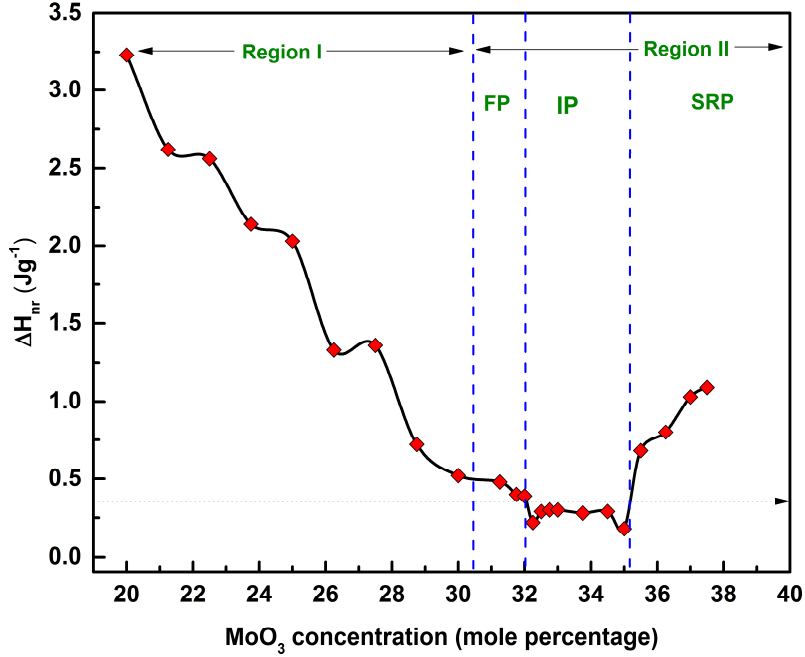

*Figure 6: Non-reversing enthalpy, $\Delta H_{nr}$ ($Jg^{-1}$) vs. $MoO_3$ concentration (mole percentage) (x)*

The heat capacity study confirms the presence of nanocluster within $20 < x \leq 30$ compositional range for samples $(AgI)_{75-x}$-$(Ag_2O)_{25}$-$(MoO_3)_x$. The gradual formation of [– Mo – (O – Ag) – $O_L$ –] bonds within the system incorporates with the decreasing profile of $\Delta H_{nr}$ with $MoO_3$ concentration, indicating a lowering value of configurational entropy difference because $\Delta S_c = S_{c\,liquid} - S_{c\,crystal} = \Delta H_{nr}/T_g$. Beyond this threshold, the glassy system becomes necessarily and sufficiently covalent to exhibit the rigidity percolation phenomena. The significance of the rigidity phases are well studied [8-14, 16-19]. In this work, we also pursue the same context of the significance. The floppy phase (FP) or the under-constrained region is found to be within $31.25 \leq x \leq 32$ range. Beyond the FP, the system starts becoming rigid. The over-constrained, stressed and rigid phase (SRP) appears to be in the range beyond $x \geq 35.5$. Between these two region, within a range $32.25 \leq x \leq 35$, a square well like global minima appears, where the value of $\Delta H_{nr}$ becomes very less and hence $\Delta S_c \sim 0$. This phase is happened to be mechanically rigid but stress free and known to be the intermediate phase. This study confirms the presence of rigidity percolation and nanocluster in the ion conducting, conditional glass forming system, $(AgI)_{75-x}$-$(Ag_2O)_{25}$-$(MoO_3)_x$, which can corroborated further with various distinct features of this system to improve the applicability and understanding. The implications of these phases, especially IP, in the context of structure and conductivity have been verified but are beyond the scope of this work.

## IV. Conclusion

In this work we have investigated the rigidity percolation phenomena in a fast ion conducting, conditional glass forming system, AgI-Ag$_2$O-MoO$_3$. The appearance of rigidity percolation in an ion conducting, conditional oxide glass system is not as apparent as in case of covalent chalcogen based glasses because of lack of covalent, CRN structure. A wide range of composition accommodates two types of glasses: increase in AgI concentration results in fragile glass formation, increase in MoO$_3$ concentration introduces covalent bonds into the system and forms strong glass. Heat capacity near T$_g$, plays an important role to identify the essentials: it's fluctuation behavior classifies the fragile and strong glasses and it's abrupt sign shift determines the threshold between these two types of glasses. Negative heat capacity in the fragile region has been corroborated with the existence of nanoclusters and a cluster (dense) -tissue (less dense) type structural model. The XPS study concludes the formation of [– Mo – (O – Ag) – O$_L$ –] bonds to give rise to the required covalent nature within the strong region. Finally, the measured values of ΔH$_{nr}$ from ADSC experiment, exhibit a well like global minima within the range $32.25 \leq x \leq 35$ which is the intermediate phase where the system is rigid but stress free and the configurational entropy difference between solid and liquid becomes very less.